\documentstyle[psfig,times]{mn}

\def\etal{et~al.}
\def\spose#1{\hbox to 0pt{#1\hss}}
\def\lta{\mathrel{\spose{\lower 3pt\hbox{$\mathchar"218$}}
     \raise 2.0pt\hbox{$\mathchar"13C$}}}
\def\gta{\mathrel{\spose{\lower 3pt\hbox{$\mathchar"218$}}
     \raise 2.0pt\hbox{$\mathchar"13E$}}}

\def\sextractor{{\sc sextractor}}
\def\iraf{{\sc iraf}}

\def\dimsum{{\sc dimsum}}

\def\kms{\,km\,s\,$^{-1}$}
\def\Ho50{$H_0 = 50$km\,s$^{-1}$\,Mpc$^{-1}$}

\title[Environments of $z \sim 1.6$ radio sources]{Red galaxy overdensities
and the varied cluster environments of powerful radio sources with $\mathbf {z
\sim 1.6}$}

\author[P.~N.~Best \etal]{P.~N.~Best,$^1$\thanks{Email:
pnb@roe.ac.uk} M. D. Lehnert$^2$, G.K. Miley$^3$,
H.J.A. R{\"o}ttgering$^3$\\   
$^1$ Institute for Astronomy, Royal Observatory Edinburgh, Blackford Hill,
Edinburgh EH9 3HJ, UK\\
$^2$ Max-Planck-Institut f{\"u}r extraterrestrische Physik, Postfach 1312, 
85741 Garching, Germany\\ 
$^3$ Sterrewacht Leiden, Postbus 9513, 2300 RA Leiden, the Netherlands\\ 
}

\pagerange{\pageref{firstpage}--\pageref{lastpage}}
\pubyear{2002}
\begin{document}
\label{firstpage}

\maketitle

\begin{abstract}
\noindent The environments of a complete subsample of 6 of the most powerful
radio--loud AGN at redshifts $z \sim 1.6$ are investigated, using deep $RJK$
imaging to depths of $R \sim 26$, $J \sim 22.4$ and $K \sim 20.6$. An excess
of galaxy counts in the K--band is seen across these fields; these surplus
galaxy counts are predominantly associated with red galaxies ($R-K \gta 4$) of
magnitudes $17.5 \lta K \lta 20.5$ found within radial distances of
$\sim$1\,Mpc of the AGN host. These are exactly (though not uniquely) the
magnitudes, colours and locations that would be expected of old passive
elliptical galaxies in cluster environments at the redshifts of these AGN.
Using both an Abell--style classification scheme and investigations of the
angular and spatial cross--correlation functions of the galaxies, the average
environment of the fields around these AGN is found to be consistent with
Abell cluster richness classes 0 and 1. The amplitude of the angular
cross-correlation function around the AGN is shown to be a strong function of
galaxy colour, and is highest when only those galaxies with the colours
expected of old elliptical galaxies at these redshifts are considered.

The images cover a relatively wide field, 5 by 5 arcmins, allowing the
distribution of the surplus galaxy counts to be investigated.  The galaxy
overdensities are found on two scales around the AGN: (i) pronounced central
concentrations on radial scales within $\sim 150$\,kpc; where present, these
are composed almost entirely of red ($R-K \gta 4$) galaxies, suggesting that
the morphology--density relation is imprinted into the centres of clusters at
a high redshift.  (ii) weaker large--scale excesses extending out to between 1
and 1.5\,Mpc radius. The presence or absence of galaxy excesses on these two
scales, however, differs greatly between the six different fields: the fields
of two AGN do show red galaxy excesses on both scales, another two fields show
only a large--scale red galaxy overdensity with no pronounced central
concentration, and one field shows only a sharp central peak of red galaxies
with no large--scale overdensity. The final field shows little evidence for an
excess on any scale; this field is associated with an unresolved radio source,
perhaps indicating that only extended radio sources probe cluster
environments.

Clearly there is a large range in both the richness and the degree of
concentration of any clustering environments around these distant AGN. The
implications of this for both cluster formation and the nature of high
redshift AGN are discussed.
\end{abstract}


\section{Introduction}

Understanding the formation of large scale structure and the related formation
and evolution of galaxies are two of the most important issues in modern--day
astronomy.  Rich clusters of galaxies at high redshifts ($z \gta 1$) provide
an opportunity to investigate both of these problems directly: first, since
clusters are the largest, most massive, collapsed structures in the Universe,
discerning the epoch and process of their assembly places strong constraints
on models of large scale structure formation (e.g. Bahcall \& Fan
1998);\nocite{bah98} second, because they contain large numbers of galaxies at
the same distance, clusters provide a unique resource for investigating galaxy
evolution through the evolution of galaxy scaling relations such as the
fundamental plane (e.g. van Dokkum et~al 1998) and colour--magnitude relations
(e.g. Stanford et~al. 1998);\nocite{dok98b,sta98} third, clusters contain the
oldest galaxies known, and these can strongly constrain the first epoch of
formation of ellipticals (cf Dunlop \etal\ 1996, Spinrad \etal\
1997).\nocite{dun96,spi97}

The identification of significant samples of clusters much above redshift one
is both challenging and inefficient using optical or X--ray selection
techniques. Existing X--ray surveys suffer from sensitivity limits (the ROSAT
Deep Cluster Survey, for example, found very few clusters above $z \sim 1$,
and none above $z=1.3$; e.g. Stanford \etal\ 2002)\nocite{sta02} whilst the
smaller field of view of the current generation of X-ray telescopes make them
relatively inefficient for wide--area surveys.  At optical wavelengths the
contrast of a cluster above the background counts is minimal above redshift
one, requiring very deep wide--area multi--colour surveys including
near--infrared wavebands to identify a significant sample of $z \gg 1$
clusters; whilst certainly feasible, this is very expensive on telescope
time. For this reason, an alternative method has frequently been adopted,
using targetted studies towards powerful AGN. At low redshifts AGN are
frequently found in relatively rich environments (e.g. Yates, Miller and
Peacock 1989; Hill \& Lilly 1991)\nocite{yat89,hil91}, and the host galaxies
of high redshift AGN are amongst the most massive galaxies known in the early
Universe (e.g. de Breuck \etal\ 2002),\nocite{bre02a} making these promising
candidates for residing in galaxy overdensities.

The environments of powerful radio sources at redshifts $z \sim 1$ have been
investigated by many authors over recent years using a variety of different
techniques, including searches for luminous extended X--ray emission
(e.g. Crawford \& Fabian 1996; Fabian 2001)\nocite{cra96b,fab01}, searches for
over-densities of galaxies in infrared colours (e.g. Dickinson 1997, Hall \&
Green 1998)\nocite{dic97a,hal98} or emission line images (e.g. McCarthy \etal\
1995)\nocite{mcc95}, investigations of the colour--magnitude relations for red
cluster ellipticals (e.g. Best 2000, hereafter B00), cross--correlation
analyses (e.g. B00; Wold et~al 2000)\nocite{bes00f,wol00} and direct
spectroscopic studies of individual sources (e.g. Dickinson 1997, Deltorn
\etal\ 1997)\nocite{dic97a,del97}.  The evidence that at least some powerful
$z \sim 1$ radio sources are located at the centres of galaxy overdensities is
overwhelming (see B00 for a review). What remains unclear is the ubiquity,
scale, and nature of these (proto?)  clusters.

At higher redshifts, radio sources have been detected out to $z=5.2$
\cite{bre99a}, and some well--studied sources have been spectroscopically
confirmed to lie in overdense environments (e.g. 1138-215 at $z=2.2$; Kurk
\etal\ 2000; Kurk \etal\ in preparation).\nocite{kur00b} An on--going VLT
large project (PI: Miley) has found order--of--magnitude overdensities of
Ly-$\alpha$ emitters around all 4 deeply studied $z>2$ radio galaxies to date
(Venemans \etal\ in preparation), including the detection of the most distant
known large structure around the $z = 4.1$ radio galaxy TN J1338-1942
\cite{ven02}. The $\sim 20$ line emitting galaxies in the structures around
each of these high redshift radio galaxies are found over regions extending to
$\gta 3$\,Mpc, and have velocity dispersions of 300--1000\kms (Venemans \etal,
private communication). Thus, it does indeed seem that powerful radio galaxies
offer unique\footnote{Of course, within a few years deeper X--ray surveys and
Sunyaev--Zel'dovich effect surveys will also find clusters at redshifts $z \gg
1$, if such clusters already contain a significant hot gas component.}
prospects for tracking galaxy clusters over cosmic time. The detection of
overdensities at these redshifts, however, is based predominantly on observing
emission--line galaxies near the radio galaxy redshift; the strong red
sequences characteristic of nearby clusters are not observed towards these $z
> 2$ overdensities, restricting the evolutionary studies of cluster
ellipticals using the fundamental plane and colour--magnitude relations from
being extended to these high redshifts.

The redshift range $1.2 \lta z \lta 2.0$ is therefore a particularly
interesting redshift range to study. This is the redshift range where cluster
studies go beyond those currently being carried out using optical and X--ray
selected clusters, whilst still being of low--enough redshift that red
sequences of cluster ellipticals are likely to exist, and can be studied in a
practical amount of observing time. Such clusters will also allow a link to be
made between the `proto'-clusters being found around $z>2$ radio galaxies, and
the rich virialised clusters seen in the nearby Universe. The elliptical
galaxy population in clusters may well be undergoing a fundamental change
within this redshift range

Hall \& Green \shortcite{hal98} have previously investigated the small--scale
environments of a sample of 31 quasars with redshifts $1 \lta z \lta 2$,
variably selected from previous surveys. They found an excess of K--band
galaxy counts on two scales around the quasars; a pronounced peak in the
galaxy counts within the central 40 arcsec radius surrounding the quasars was
accompanied by a uniform galaxy overdensity across their entire fields (to
about $\sim 1.5$ arcmins radius). Comparing R and K--band data they showed
that the excess counts correspond to redder galaxies than the typical field
population, with a significant overdensity of extremely red objects. This is
indicative that a red elliptical sequence may be present in these fields. Hall
\etal\ \shortcite{hal01} investigated the physical extent of the large--scale
overdensity in the fields of two of these AGN, and found a net excess of
galaxy counts out to 140 arcsec ($\sim 1.3$\,Mpc) in one of the two fields.

We have embarked upon a project to investigate in detail the environments
of a complete sample of the most luminous radio sources (radio galaxies
and quasars) within the narrow redshift range $1.44 \le z \le 1.7$. The
aim of this project is to use deep, wide--area multi--colour imaging to
study in detail any galaxy overdensity, in particular the nature of the
excess galaxies, the presence of a red sequence of cluster ellipticals,
and the radial extent of the overdensity. By studying a complete AGN
sample it will be possible to investigate the variety of environments
found, and whether the environment of a radio--loud AGN is in any way
correlated with the radio source properties. The candidate cluster
galaxies in the richest overdensities can then be followed--up using
multi--object spectroscopy in order to extend the galaxy evolution
studies.

This paper provides the initial results of this project, focussing on the
imaging data and results from the first 6 clusters in the sample.
Specifically, the lay-out of the paper is as follows. In Section~\ref{obssect}
the sample selection, observations, and data reduction are described. The
resulting multi--colour images of each field are provided and described in
Section~\ref{fields}, and galaxy catalogues are constructed from these. The
galaxy number counts and the colour distribution of the galaxies are
investigated in Section~\ref{galcnts}, and a cross--correlation analysis
carried out. 
The properties of the different fields are related to the
properties of their AGN in Section~\ref{envcomp}. The results are discussed
and conclusions drawn in Section~\ref{discuss}. Throughout the paper, the
values adopted for the cosmological parameters are $\Omega_{\rm M} = 0.3$,
$\Omega_{\Lambda} = 0.7$ and $H_0 = 65$\,km\,s$^{-1}$Mpc$^{-1}$.

\section{Observations}
\label{obssect}

\subsection{Sample Selection and Observations}

The radio sources were drawn from the equatorial sample of 178 powerful
radio sources defined by Best, R{\"o}ttgering and Lehnert (1999; the BRL
sample). This sample was designed to be equivalent to the northern 3CR
sample, and was drawn from the Molonglo radio survey at 408\,MHz
\cite{lar81}, according to three criteria: (i) the sources must have
declinations $-30^{\circ} < \delta < +10^{\circ}$; (ii) they must lie away
from the galactic plane, $|b| > 10$; (iii) they must be brighter than
$S_{\rm 408 MHz} = 5$\,Jy (roughly equivalent to the definition of the 3CR
sample). Spectroscopic redshifts are available for 177 of these 178
sources \cite{bes99e,bes00a}.

From the BRL sample, the 9 radio sources with redshifts $1.44 \le z \le
1.7$ and galactic latitudes $|b| > 25$ were selected. This sample
comprised 4 radio galaxies and 5 radio loud quasars. Of these, the
environments of 6 radio sources have been studied to date, and are
presented in this paper. These 6 radio sources, limited by the constraints
of telescope time, correspond to all of those with $15 < {\rm RA} < 02$
and therefore constitute a complete subsample of the most powerful radio
sources at these redshifts. 

The aim of the current observations was to deeply image these fields in the
$R$, $J$ and $K$ wavebands. This filter combination was chosen to provide
maximal sensitivity to old cluster ellipticals: the strongest continuum
features of such ellipticals is the 4000\AA\ break, which at $z \sim 1.5$
falls between the $R$ and $J$ bands, making this colour very sensitive to old
cluster galaxies. The addition of the $K$--band data provides a long red
colour baseline to sample the old stellar populations of these ellipticals,
whilst also facilitating identification of bluer cluster galaxies with
on-going star formation (and hence less pronounced 4000\AA\ breaks), and
helping to filter out non--cluster objects with very red $R-J$ colours, such
as dusty starbursting galaxies. At $z \sim 1.5$ passively evolving old
ellipticals would have colours of $J-K \sim 1.8$ and $R-J \sim 3.7$, and
brightest cluster galaxies at these redshifts are expected to have magnitudes
of $K \sim 17.5$ to 18 (cf. those of radio galaxies at these redshifts,
e.g. Best \etal\ 1998)\nocite{bes98d}. The goal of detecting elliptical
galaxies in all colours to $\sim 3$ magnitudes fainter than this set the
required depths of the different observations. Full details of the
observations are provided in Table~\ref{obsdettab}.

The six fields were observed in the $J$ and $K$--bands using the infrared
imager SOFI on the ESO New Technology Telescope (NTT) during September 2000
and August 2001. SOFI has a 1024 by 1024 HgCdTe array with 0.29 arcsec pixels,
providing a field of view of 5 by 5 arcmins. Images were taken using a
jittering technique, with the pointing centre moved randomly within a 40
arcsec square box between each exposure. At each pointing position, 1--minute
of data was taken, split into six 10s co-adds in the K--band and three 20s
co-adds in the J--band. In the $K$--band the $K_s$ filter was used to reduce
the sky background; the $K_s$ magnitude is related to the $K$-magnitude by $K
- K_s \approx -0.005(J-K)$, and all quoted magnitudes have been converted to
the $K$--band magnitude. Between 1.7 and 2.5 hours of data were obtained for
each source (see Table~\ref{obsdettab} for details), reaching a `limiting
magnitude' of $K \sim 20.6$; this value corresponds to the 50\% completeness
limit for galaxies with sizes typical of those at redshifts $z \sim 1.6$. In
the $J$--band the total exposure time varied between 1 and 2 hours, reaching a
limiting depth of $J \sim 22.4$.

For 5 of the 6 fields (2128$-$208 was randomly excluded due to telescope time
constraints), $R$-band data for the field were taken through the Bessel $R$
filter using SUSI-2 on the NTT. SUSI-2 comprises two 2048 by 4096 pixel EEV
CCDs, with a pixel scale of 0.08 arcsec, which combined provide a 5.5 by 5.5
arcminute field of view. The two CCDs are separated by about 10 arcseconds on
the sky; the total exposure time was therefore separated into 5-min blocks,
with the pointing position moved through a 9-point pattern of offsets [(0,0),
(10,10), (20,-10), (30,0), (40,10), (-40,-10), (-30,0), (-20,10), (-10,-10)
arcsec] in order to attain equal sensitivity over the entire SOFI field. This
pattern was repeated until between 3.75 and 6 hours of data had been obtained
on each field, reaching a limiting depth of $R \sim 25.9$.

Standard stars were observed approximately every 2 hours throughout the nights
to allow the images to be photometrically calibrated. For the SUSI-2 images,
once each night the image was rotated by 180 degrees in order to place the
standard stars on the opposite CCD; no significant differences in calibration
were apparent between the two CCDs. All of the nights were photometric, and
the instrument zero-points varied by less than 0.05 magnitudes within each of
the Sept 2000 and Aug 2001 runs, although the $R$--band zero point changed by
0.1 magnitudes between the two runs.
 
\begin{table*}
\caption{\label{obsdettab} Details of the observations.}
\begin{tabular}{lccccllccc}
~~~~~Source& RA & Dec & z &Waveband&\hspace*{6mm} Obs.& Telescope & Exp.& Mean & Limiting \\
             &\multicolumn{2}{c}{J2000}&&     &\hspace*{5mm} Dates& ~~~~~~\&  & Time&Seeing& Magnitude$^*$\\
             &&&&       &                 & Instrument&[mins]&[arcsec]&   \\       
\\
MRC 0000-177 & 00 03 21.93 & $-$17 27 11.9 & 1.47 & $K_s$ & 2001 Aug 10,12  & NTT-SOFI  & 138 & 0.98 & 20.4 \\
             &             &               &      &   J   & 2001 Aug 11,12  & NTT-SOFI  & 120 & 1.01 & 22.3 \\
             &             &               &      &   R   & 2001 Aug 12     & NTT-SUSI2 & 270 & 1.00 & 25.8 \\
MRC 0016-129 & 00 18 51.36 & $-$12 42 34.4 & 1.59 & $K_s$ & 2000 Sep 23     & NTT-SOFI  & 103 & 0.62 & 20.7 \\
             &             &               &      &   J   & 2000 Sep 23,25  & NTT-SOFI  &  60 & 0.80 & 22.3 \\
             &             &               &      &   R   &2000 Sep 20,24,25& NTT-SUSI2 & 225 & 0.85 & 25.8 \\
MRC 0139-273 & 01 41 27.31 & $-$27 06 11.1 & 1.44 & $K_s$ & 2000 Sep 23,25  & NTT-SOFI  & 135 & 0.64 & 20.6 \\
             &             &               &      &   J   & 2000 Sep 23,25  & NTT-SOFI  & 105 & 0.74 & 22.6 \\
             &             &               &      &   R   &2000 Sep 20,24,25& NTT-SUSI2 & 360 & 0.74 & 26.1 \\
MRC 1524-136 & 15 26 59.45 & $-$13 51 00.2 & 1.69 & $K_s$ & 2001 Aug 10,11  & NTT-SOFI  & 135 & 0.80 & 20.5 \\
             &             &               &      &   J   & 2001 Aug 10     & NTT-SOFI  & 105 & 0.90 & 22.4 \\
             &             &               &      &   R   & 2001 Aug 11,12  & NTT-SUSI2 & 255 & 1.15 & 25.5 \\
MRC 2025-155 & 20 28 07.68 & $-$15 21 20.9 & 1.50 & $K_s$ & 2000 Sep 20,23  & NTT-SOFI  & 114 & 0.75 & 20.4 \\
             &             &               &      &   J   & 2000 Sep 23     & NTT-SOFI  &  91 & 0.90 & 22.3 \\
             &             &               &      &   R   & 2000 Sep 24,25  & NTT-SUSI2 & 260 & 0.65 & 25.9 \\
MRC 2128-208 & 21 31 01.49 & $-$20 36 56.4 & 1.62 & $K_s$ & 2001 Aug 10     & NTT-SOFI  & 150 & 0.77 & 20.6 \\
             &             &               &      &   J   & 2001 Aug 10     & NTT-SOFI  & 105 & 1.00 & 22.5 \\
\\
\end{tabular}
\\
$^*$ The `limiting magnitude' quoted is the approximately 50\%
completeness limit for galaxies with physical sizes typical of those in a
$z \sim 1.6$ cluster. 
\end{table*}

\subsection{Data reduction}

The infrared observations were reduced in the standard manner, using the
\dimsum\footnote{\dimsum\ is the `Deep Infrared Mosaicing Software' package
developed by Eisenhardt, Dickinson, Stanford and Ward.} package within
\iraf. After subtraction of a dark frame, and removal of known bad pixels, a
median of the images (after masking of bright objects) was constructed to obtain
an accurate sky flat--field. The flat--fielded images were then sky subtracted,
cosmic ray events were removed, and then the individual frames were block
replicated by a factor of 2 in each dimension to allow more accurate alignment,
resulting in a final image scale of 0.144 arcsec per pixel. The frames were
accurately registered using the peak positions of several bright unresolved
objects visible on all images, and were combined. Because of the long exposure
times, the seeing sometimes varied throughout the dataset, and so during the
combination process the images were weighted according to their seeing in order
to provide a final image with the optimal combination of seeing and
sensitivity. Stars from the US Naval Observatory catalogue A2.0 \cite{mon98}
were used to provide an astrometric solution for the images. The absolute
astrometry should be accurate to better than 0.5 arcsec. The fields were
photometrically calibrated onto the standard Vega photometric system.

The $R$--band images were reduced within \iraf. The images were bias
subtracted, and then flat--fielded using a flat--field constructed by from the
median of the various images (after masking bright objects). An illumination
correction was then applied to remove large--scale variations in the
flat--fielded images. The images were registered using the peak positions of
several bright unresolved objects, and then they were combined, weighted
according to their seeing as for the $J$ and $K$ band data. Cosmic rays were
rejected during the combination process. The resultant image was then
astrometrically transformed using a 5 parameter fit (RA and Dec of the central
pixel, pixel scale in x and y directions, and rotation angle) to align it with
the infrared data, converting it to 0.144 arcsec pixels whilst conserving the
object magnitudes. The relative astrometry of the IR and optical frames is
accurate to $< 0.1$ arcsec.

\section{The radio source fields}
\label{fields}

K-band images of the six fields are shown in Figures~\ref{0000+0016kfig}
to~\ref{2025+2128kfig}, with the radio source host galaxy marked by a
cross in each case. Three--colour images of the 100 by 100 arcsec region
around the radio source host in each field are provided in
Figures~\ref{0000+0016colfig} to~\ref{2025+2128colfig}. A brief discussion
of each individual field follows.

\noindent {\bf 0000--177 ($\mathbf{z=1.47}$):} This radio--loud quasar lies
in a relatively sparse region of sky. A faint object is seen 1-2 arcsec to the
north--east of the quasar host, and a further seven objects with $R-K > 4$ lie
within 30 arcsec ($\approx 275$\,kpc at $z \sim 1.5$), indicating a possible
compact group of galaxies at the quasar redshift. There is no obvious
large--scale galaxy overdensity.

\noindent {\bf 0016--129 ($\mathbf{z=1.59}$):} This radio galaxy is the
westernmost of a string of three luminous red galaxies in the centre of this
field, and shows extended blue emission along the radio axis direction,
characteristic of the alignment effect in high redshift radio galaxies
(cf. McCarthy \etal\ 1987, Chambers \etal\ 1987).\nocite{mcc87,cha87} The
redder galaxies in the field are more concentrated towards the central regions
of the field, within 1--2 arcmins of the AGN.

\noindent {\bf 0139--273 ($\mathbf{z=1.44}$):} This is the richest field of
those imaged in the current observations, and contains a large number of very
red galaxies. These galaxies are found predominantly within the inner regions
of the field, although no red galaxies are seen close to the AGN.  Very
luminous elongated blue emission extends from the radio galaxy host, along the
direction of the radio source axis.

\noindent {\bf 1524--136 ($\mathbf{z=1.69}$):} This radio loud quasar is the
highest redshift AGN in the current sample, and its surrounding field is
relatively rich in the K--band. A moderate excess of red galaxies is seen on
scales out to about 120 arcsec ($\sim$ 1\,Mpc). No sharp central peak (such as
that seen for 2025$-$155) is seen, suggesting that any cluster environment
may still be forming.

\noindent {\bf 2025--155 ($\mathbf{z=1.50}$):} A very compact group of
several red galaxies is found within 15--20 arcsec (150\,kpc) radius of
this radio--loud quasar. All of these objects have colours consistent with
old passive ellipticals at the radio source redshift. A significant
overdensity of red galaxies is found extending over the entire 5 arcmin
field, although predominantly within the inner regions.

\noindent {\bf 2128--208 ($\mathbf{z=1.62}$):} This field does not contain
a particular excess of K-band galaxies and, except for a companion object
to the quasar, no other galaxy is found within 15 arcsec of the AGN
host. With only J and K--band data available, there is little information
about the colour distribution of the galaxies across the field.

It is clear from this direct examination of the different images that a wide
variety of environments in found amongst the sample. At one extreme, the field
of 2128$-$208 shows little evidence for any net overdensity of red galaxies on
either small ($\lta 150$\,kpc) or large ($\sim$\,Mpc) scales, whilst in
contrast 2025$-$155 has a very rich environment on small scales coupled with
an apparent excess of red galaxies across the entire field. Between these
extremes, the other fields show evidence for structure on differing scales: a
small scale excess of red galaxies is found around 0000$-$177, but not much
large scale structure; the 0016$-$129 field has three central red galaxies and
a weak larger--scale excess; above average numbers of red galaxiess are seen
on both moderate and large scales around 1524$-$136; and the field of
0139$-$273 shows a strong excess of red galaxies on larger scales, but with
few red objects close to the AGN. In the discussions of the statistical
properties of the fields that follow, it is important to bear in mind that
global results quoted correspond to the average values across 6 rather
different environments.

\begin{figure*}
%\centerline{
%\psfig{file=0000kfin.ps,angle=90,width=12cm,clip=}
%}
%\centerline{
%\psfig{file=0016kfin.ps,angle=90,width=12cm,clip=}
%}
\caption{\label{0000+0016kfig} [See attached jpg files] {\it Top:} The K-band image of the field of
0000$-$177 ($z=1.47$). {\it Bottom:} The K-band image of the field of
0016$-$129 ($z=1.59$).}
\end{figure*}

\begin{figure*}
%\centerline{
%\psfig{file=0139kfin.ps,angle=90,width=12cm,clip=}
%}	    	
%\centerline{	
%\psfig{file=1524kfin.ps,angle=90,width=12cm,clip=}
%}
\caption{\label{0139+1524kfig} [See attached jpg files] {\it Top:} The K-band image of the field of
0139$-$273 ($z=1.44$). {\it Bottom:} The K-band image of the field of
1524$-$136 ($z=1.69$).}
\end{figure*}

\begin{figure*}
%\centerline{
%\psfig{file=2025kfin.ps,angle=90,width=12cm,clip=}
%}	    	
%\centerline{	
%\psfig{file=2128kfin.ps,angle=90,width=12cm,clip=}
%}
\caption{\label{2025+2128kfig} [See attached jpg files] {\it Top:} The K-band image of the field of
2025$-$155 ($z=1.50$). {\it Bottom:} The K-band image of the field of
2128$-$208 ($z=1.62$).}
\end{figure*}

\begin{figure*}
%\centerline{
%\psfig{file=0000rjk.ps,angle=-90,width=11cm,clip=}
%\vspace*{1cm}	   
%}	    	   
%\centerline{	   
%\psfig{file=0016rjk.ps,angle=-90,width=11cm,clip=}
%}
\caption{\label{0000+0016colfig} [See attached jpg files] {\it Top:} An R,J,K colour image of the
central 100 $\times$ 100 arcsec field around the $z=1.47$ quasar
0000$-$177. {\it Bottom:} An R,J,K colour image of the central 100
$\times$ 100 arcsec field around the $z=1.59$ radio galaxy 0016$-$129.}
\end{figure*}

\begin{figure*}
%\centerline{
%\psfig{file=0139rjk.ps,angle=-90,width=11cm,clip=}
%\vspace*{1cm}	   
%}	    	   
%\centerline{	   
%\psfig{file=1524rjk.ps,angle=-90,width=11cm,clip=}
%}
\caption{\label{0139+1524colfig} [See attached jpg files] {\it Top:} An R,J,K colour image of the
central 100 $\times$ 100 arcsec field around the $z=1.44$ quasar
0139$-$273. {\it Bottom:} An R,J,K colour image of the central 100
$\times$ 100 arcsec field around the $z=1.69$ radio galaxy 1524$-$136.}
\end{figure*}

\begin{figure*}
%\centerline{
%\psfig{file=2025rjk.ps,angle=-90,width=11cm,clip=}
%\vspace*{1cm}	   
%}	    	   
%\centerline{	   
%\psfig{file=2128rjk.ps,angle=-90,width=11cm,clip=}
%}
\caption{\label{2025+2128colfig} [See attached jpg files] {\it Top:} An R,J,K
colour image of the central 100 $\times$ 100 arcsec field around the $z=1.50$
quasar 2025$-$155. {\it Bottom:} A J,K colour image of the central 100
$\times$ 100 arcsec field around the $z=1.62$ quasar 2128$-$208.}
\end{figure*}

\subsection{The source catalogues}
\label{sextract}

Source extraction and catalogue construction followed the method described
in detail by B00, and the reader is referred to that paper for full
details. This method is summarised briefly below.

Throughout these analyses, the K--band data were used as the primary
dataset from which the catalogue is defined. Source detection and
photometry were carried out on the K--band frames using \sextractor\
version 2.2.1 \cite{ber96}, using the output exposure map produced by
\dimsum\ as a weight map for \sextractor\ to compensate for the varying
noise levels associated with the jittering procedure. The source extraction
parameters were set such that, to be detected, an object must have a flux in
excess of 1.5 times the local background noise level over at least $N$
connected pixels, where $N$ was varied a little according to the seeing
conditions, but in general was about 15 (equivalent to 4 pixels prior to the
block replication during the mosaicing procedure). A search for negative holes
using the same extraction parameters (cf B00) found only one negative feature
over all 6 frames brighter than the 50\% completeness limit, implying that
essentially all of the extracted objects down to that level are real. The
\sextractor\ catalogues were examined carefully, and obvious problems, such as
objects associated with the diffraction spikes of bright stars, were
corrected. Those objects within the central 2000 by 2000 pixels (290 by 290
arcsec) of each image were retained for the subsequent analysis; more distant
objects lie in the regions of the image where the noise levels are higher due
to the jittering process, and were discarded; at the edges of the retained
regions of the fields, the noise level is only $\sim 10$\% higher than in the
field centre. Small regions of sky (of 10, 20 and 10 arcsecond radius) were
also discarded around the three bright stars in the images of 0016$-$129,
1524$-$136 and 2025$-$155. The six fields combined provide a sky area of 140
square arcminutes.

\sextractor's {\sc mag\_best} estimator was used to determine the
magnitudes of the extracted sources; this yields an estimate for the
`total' magnitude using Kron's \shortcite{kro80} first--moment algorithm,
except if there is a nearby companion which may bias the total magnitude
estimate by more than 10\% in which case a corrected isophotal magnitude
is used instead. \sextractor\ isophotal apertures smaller than 1 arcsecond
were replaced by circular 1 arcsecond apertures. The determined magnitudes
were corrected for galactic extinction using the $E(B-V)$ values provided
in the NASA\,/\,IPAC Extragalactic Database (NED). The accuracy of these
total magnitudes and the completeness level of the source extraction were
investigated as a function of position on each image through a
Monte--Carlo simulation, as fully described by B00. Briefly, a total of
10000 stellar objects and 25000 galaxies (see B00 for the prescription by
which these were constructed) were added, 25 at a time, to each image with
a magnitude between 15 and 22.5, and were then re-extracted using
\sextractor\ with the same input parameters as for the original source
extraction, to see if the added objects were detected. For each recovered
source, the difference between the total magnitude measured by
\sextractor\ and the input total magnitude was determined.

\begin{figure}
\centerline{
\psfig{file=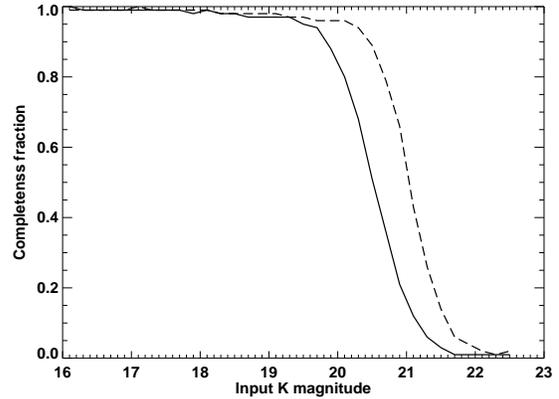,angle=90,width=7.8cm,clip=}
}
\caption{\label{compplot} The completeness fraction for both stellar
(dashed line) and galactic (solid line) model objects as a function of
input magnitude, averaged over all 6 fields.}
\end{figure}

\begin{figure}
\centerline{
\psfig{file=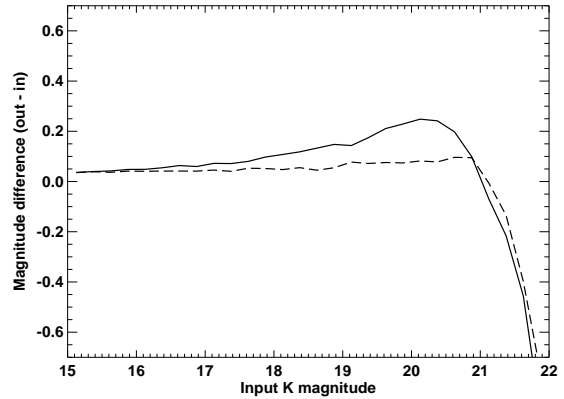,angle=90,width=7.8cm,clip=}
}					     
\centerline{				     
\psfig{file=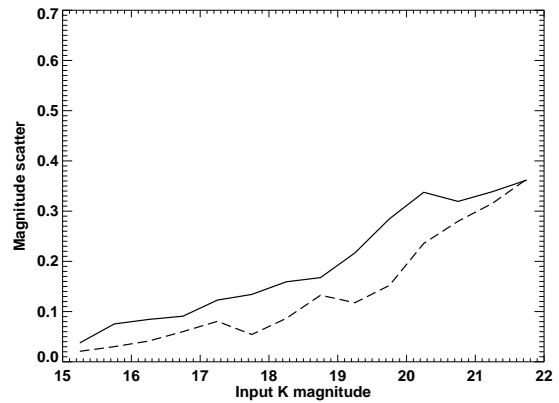,angle=90,width=7.8cm,clip=}
}
\caption{\label{inmagplots} The mean (upper plot) and rms scatter (lower
plot) of the difference between the measured magnitudes and input
magnitudes for the model stars (dashed line) and galaxies (solid line).}
\end{figure}

From these results, the mean completeness fraction was determined as a
function of magnitude for both stars and galaxies over the entire set of
images. The mean 50\% completeness limits are $K=20.55$ for galaxies and
$K=21.05$ for stars (see Figure~\ref{compplot}); the galactic completeness
limits for each individual field are provided in Table~\ref{obsdettab}.
The mean difference between the input and measured magnitudes, and the
scatter of the measured magnitudes around this mean, are shown as
functions of input magnitude for both model stars and galaxies in
Figure~\ref{inmagplots}. This difference was used to correct the observed
magnitudes and obtain true total magnitudes.

The $J$ and $R$ band magnitudes for the catalogue sources were then measured
by running \sextractor\ in its double image mode, where one image (the $K$
band) is used to detect the objects and the fluxes and magnitudes are then
measured from the second ($J$ or $R$ band) image through the same
aperture. Those objects whose $J$ or $R$ band fluxes were measured to have a
significance of less than 3$\sigma$, where $\sigma$ is the uncertainty on the
flux measurement provided by \sextractor, were replaced by 3$\sigma$ upper
limits. The \sextractor\ flux error estimate includes the uncertainty due to
the Poisson nature of the detected counts and that from the standard deviation
of the background counts. An additional source of flux error arises from the
uncertainty in the subtraction of the background count level as a function of
position across the image. This value was estimated as the product of the area
of the extraction aperture and the rms variation of the subtracted background
flux across blank regions of the image. This background subtraction error
estimate was combined in quadrature with the flux error given by \sextractor\
to determine the uncertainties on the magnitudes of the extracted objects in
all three colours.

\subsection{Star--galaxy separation}
\label{stargalsep}

\begin{figure}
\centerline{
\psfig{file=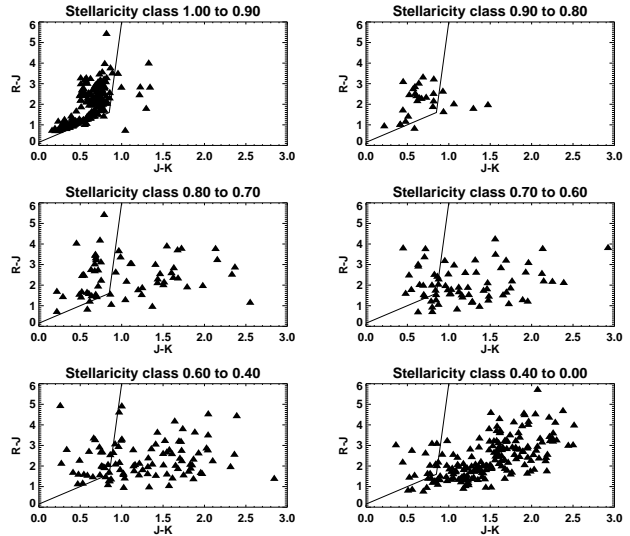,angle=90,width=8.2cm,clip=}
}
\caption{\label{colsepfig} A set of $R-K$ versus $J-K$ colour--colour
plots for objects in the field of 2025-155, separated by the stellaricity
index output by \sextractor. The solid lines define a region of colours
separating the majority of the stellar locus from typical galactic
colours. All objects with stellaricity indices greater than 0.8 were
classified as stars, together with those objects with stellaricities
between 0.5 and 0.8 lying to the upper left of the two lines.}
\end{figure}

\sextractor\ provides a `stellaricity index' for each object, where in the
ideal case a galaxy has a stellaricity index of 0.0 and a star has 1.0; low
signal--to--noise, and galaxies more compact than the seeing, lead to an
overlap in the calculated stellaricity indices for the two types of object at
the faintest magnitudes, and hence uncertain star--galaxy separation.
Stellaricity indices were taken from the K--band data. At magnitudes $K \lta
18$, star--galaxy separation could be carried out directly from these
stellaricity values taking values above 0.8 (the generally adopted cut-off) to
correspond to stars. At fainter magnitudes, separation is less certain, but
was aided by the addition of the multi--colour information; by plotting the
R-K versus J-K colour--colour diagram as a function of stellaricity index, a
region of colour space typically occupied by stars was defined (see
Figure~\ref{colsepfig}), and all objects in this colour space region with
stellaricities above 0.5 were also classified as stars. This classification is
somewhat conservative, and a small proportion of stars may be included along
with the galaxies, but this was considered preferable to falsely classifying
galaxies as stars.

\begin{figure}
\centerline{
\psfig{file=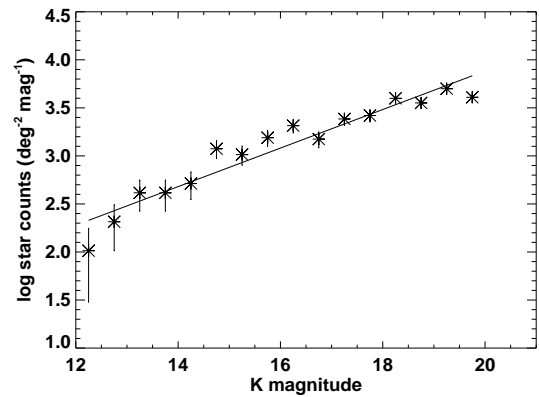,angle=90,width=7.8cm,clip=}
}
\caption{\label{starcnts} The differential star number counts in the
K--band, in $\Delta K = 0.5$ magnitude intervals, fit using a single power
law distribution.}
\end{figure}

Figure~\ref{starcnts} shows the total star counts as a function of
magnitude derived from all of the frames, fit with the function $\log N(K)
= 0.201 K - 0.128$. This relatively good fit to the differential star
counts using a single power--law distribution demonstrates that the
star--galaxy separation is working well to at least $K \sim 20$. All
objects classified as stars were removed from further analysis.

\section{Galaxy Counts and Colours}
\label{galcnts}

\subsection{Galaxy counts}
\label{subgalcnts}

\begin{figure}
\centerline{
\psfig{file=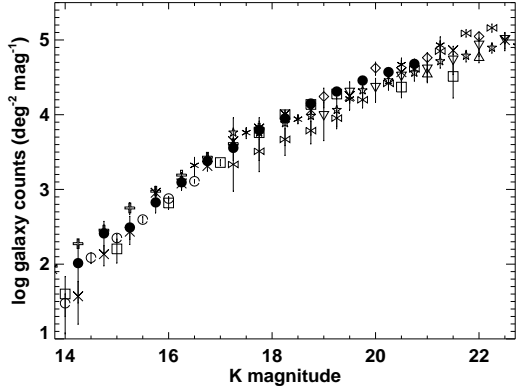,angle=90,width=7.8cm,clip=}
}
\caption{\label{galcntsplot} The K--band galaxy number counts per
magnitude per square degree. The solid symbols represent the data
presented in this paper and tabulated in Table~\ref{galcntstab}. The open
symbols represent data from the literature: squares are from Minezaki
\etal\ (1998), triangles from Djorgovski \etal\ (1995), diamonds from
Moustakas \etal\ (1997), circles from Szokoly \etal\ (1998), x's from
Gardner, Cowie \& Wainscoat (1993), asterisks from McLeod \etal\ (1995),
inverted triangles from Bershady, Lowenthal \& Koo (1998), stars from
Totani \etal\ (2001), crosses from V{\"a}is{\"a}nen \etal\ (2000), and
bow-ties ($\bowtie$) from Saracco \etal\ (1999).}
\end{figure}

The galaxy counts were determined by once again following the procedure of
B00. In brief, the measured magnitudes of the galaxies were converted to
true `total' magnitudes using the offsets determined from the simulations
above, and binned in 0.5 magnitude bins over the magnitude range $14 < K <
21$. These raw galaxy counts, $n_{\rm raw}$, were then corrected for
completeness and for magnitude biasing effects (again as calculated from
the results of the simulations) to produce corrected galaxy counts
($n_{\rm c}$).  Scaling by the observed sky area produced final counts per
magnitude per square degree, $N_{\rm c}$. In calculating the uncertainty
on this value, in addition to the Poissonian error on the raw galaxy
counts, an extra error term was included to account for the uncertainty of
the completeness correction procedure; this was estimated at 20\% of the
number of galaxies added in the correction. The resulting galaxy counts
are tabulated in Table~\ref{galcntstab} and are plotted in
Figure~\ref{galcntsplot} compared with counts from various K--band
blank--field surveys
\cite{gar93,mcl95,djo95,mou97,min98,szo98,ber98,sar99,vai00,tot01}.

The derived counts are fully consistent with the literature counts at
magnitudes $K \lta 17.5$.  This magnitude corresponds to the expected
magnitude of brightest cluster galaxies at $z \sim 1.5$. Fainter than this
magnitude an surplus of galaxies is seen; the galaxy counts still lie within
the scatter of the other observations, that scatter being largely dominated by
cosmic variance (cf Daddi \etal\ 2000a)\nocite{dad00a}, but the counts in every
magnitude bin from $17.5 < K < 20.5$ exceed the literature average by at least
1$\sigma$. 

\nocite{min98,djo95,mou97,szo98,gar93,mcl95,ber98,tot01,vai00,sar99}

\begin{table}
\caption{\label{galcntstab} K--band galaxy counts as a function of
magnitude, over the combined 140 square arcminutes. The columns give the raw
($n_{\rm raw}$) and corrected ($n_{\rm c}$) counts, the corrected counts
per square degree per unit magnitude ($N_{\rm c}$), the error on this, and
the mean literature counts ($N_{\rm lit}$).}
\begin{tabular}{crrrrr}
$K$&$n_{\rm raw}$&$n_{\rm c}$ &$N_{\rm c}$$^*$ &$\delta N_{\rm c}$$^*$
&$N_{\rm lit}$$^*$ \\
\multicolumn{2}{l}{All fields} \\
14.0 -- 14.5  &    2  &   2.0  &   103 &   72 &    86  \\
14.5 -- 15.0  &    5  &   5.0  &   257 &  115 &   175  \\
15.0 -- 15.5  &    6  &   6.0  &   309 &  126 &   340  \\
15.5 -- 16.0  &   13  &  13.0  &   670 &  185 &   710  \\
16.0 -- 16.5  &   24  &  24.2  &  1250 &  250 &  1160  \\
16.5 -- 17.0  &   46  &  46.5  &  2400 &  350 &  2280  \\
17.0 -- 17.5  &   69  &  69.7  &  3590 &  440 &  4010  \\
17.5 -- 18.0  &  118  & 120.8  &  6230 &  580 &  5530  \\
18.0 -- 18.5  &  170  & 172.8  &  8920 &  690 &  8260  \\
18.5 -- 19.0  &  254  & 263.5  & 13600 &  860 & 10500  \\
19.0 -- 19.5  &  361  & 375.3  & 19400 & 1030 & 15300  \\
19.5 -- 20.0  &  476  & 522.9  & 27000 & 1350 & 23100  \\
20.0 -- 20.5  &  472  & 685.1  & 35300 & 3500 & 31900  \\
20.5 -- 21.0  &  292  & 927.8  & 47900 & 7400 & 41800  \\
\\
\multicolumn{6}{l}{$^*$: counts per magnitude per square degree.}
\end{tabular}
\end{table}

\begin{table}
\caption{\label{cntsind} Properties of the galaxies in the individual
fields. The columns give: (1) the radio source field; (2) the K--band galaxy
counts (in galaxies per square degree) between $17.5 < K < 20.5$, and the
uncertainty of this value --- this uncertainty includes both the Poissonian
error and a cosmic variance term; (3) the deviation of the number counts from
the mean literature counts; (4 and 5) the mean and median $R-K$ colours of
these galaxies; (6) the Abell cluster strength estimator $N_{\rm 0.5}$, and
its error. The bottom two rows of the table give the values average over all
of the fields, and blank--field expectations for the number counts and average
colours as drawn from the literature (see text for details).}
\begin{tabular}{cccccc}
Field & $N_{\rm c}$ & Excess & \multicolumn{2}{c}{$R-K$} & ~$N_{\rm 0.5}$ \\
            & [deg$^{-2}$]     &              &Mean  &Median&             \\
   (1)      &       (2)        &    (3)       & (4)  & (5)  &     ~(6)    \\
0000$-$177  & $42200 \pm 5100$ & $-1.0\sigma$ & 4.19 & 4.19 &  $-3 \pm 5$ \\ 
0016$-$129  & $55400 \pm 5800$ & $+1.4\sigma$ & 4.09 & 4.05 & ~~$4 \pm 6$ \\
0139$-$273  & $64200 \pm 6200$ & $+2.7\sigma$ & 3.98 & 3.93 & ~$18 \pm 8$ \\
1524$-$136  & $59600 \pm 6100$ & $+2.0\sigma$ & 4.07 & 3.85 & ~$14 \pm 7$ \\
2025$-$155  & $58700 \pm 5900$ & $+1.9\sigma$ & 4.05 & 3.89 & ~$11 \pm 7$ \\
2128$-$208  & $50400 \pm 5600$ & $+0.6\sigma$ & ---  & ---  & ~~$3 \pm 6$ \\
Combined    & $55000 \pm 2400$ & $+3.1\sigma$ & 4.08 & 3.95 & ~~$8 \pm 3$ \\ 
Literature  & $47500 \pm 2000$ &   ---        & 3.87 & 3.78 &     ---     \\
\end{tabular}
\end{table}

In Table~\ref{cntsind} the total galaxy counts in the magnitude range $17.5 <
K < 20.5$ are provided for each individual field, and compared to the mean
value of the blank--field surveys. To carry out such a comparison it is
important to bear in mind that the uncertainty on the counts in the AGN fields
will be significantly larger than solely the Poissonian error, due to cosmic
variance and the effects of galaxy clustering. The true uncertainty on the
counts can be estimated from the equation $\sigma_{\rm true}^2 = N(1 + NAC)$,
where $N$ is the total counts, $A$ is the angular cross--correlation
amplitude, and $C$ is the integral constraint which depends upon the size and
geometry of the field. As derived in Section~\ref{crosscorsec}, for these
fields the values of $A$ and $C$ are 0.0005 and 16.7 respectively. Hence, for
$N \sim 300$ galaxies detected in each field, the uncertainty on the counts is
approximately double that which would be expected from Poissonian statistics
alone. The errors tabulated in Table~\ref{cntsind} are these true errors,
including this contribution of cosmic variance.  For the literature counts,
given that these are derived by combining many independent surveys over
different areas of sky, the combination should average out the cosmic variance
effects to a large extent.

It is noteworthy that five of the six fields have counts above the literature
average, with three of the fields being $\gta 2 \sigma$ above. Combining the
fields, the cumulative counts over this magnitude range are 3$\sigma$ above
the average of blank--field surveys; it should be emphasised that these quoted
significance levels including the effects of cosmic variance, and are twice
this size considering Poisson statistics alone. Indeed, since the cosmic
variance is caused by the clustering of galaxies, fields which contain
clusters of galaxies, like these are proposed to, are in fact the type of
field that contributes significantly to this cosmic variance, and so the
Poissonian estimate may be more appropriate.

The cumulative excess of galaxy counts corresponds to about 290 ($\pm 100$)
additional galaxies, that is, of order 50 galaxies per field. Although it is
possible that some fraction of these correspond to faint stars that were
misclassified as galaxies (cf. the fall in star counts in the final bin of
Figure~\ref{starcnts}), it is clear that these fields are, on average, richer
than average environments. For comparison, in Abell's original cluster
definition \cite{abe58} he considered the excess counts of galaxies within a
radius of 1.5$h^{-1}$\,Mpc of the cluster centre, with magnitudes between
$m_3$ and $m_3 + 2$, where $m_3$ is the magnitude of the third ranked cluster
member: an excess of 30 to 49 galaxies corresponded to Abell Class 0 clusters,
and an excess of 50 to 79 galaxies was Abell Class 1. Assuming that the radio
host galaxies are of order a magnitude brighter than a third--ranked cluster
member\footnote{If the radio galaxies were dramatically brighter in the K-band
than other cluster members, as is the case for some cD galaxies in nearby
clusters, then that would require the fields to be very rich, probably
implausibly so.}  the average excess number counts correspond to moderate
richness clusters.

To obtain a better indication of the cluster richnesses, the parameter
$N_{0.5}$ defined by Hill and Lilly \shortcite{hil91} can be used. This is an
Abell--type measurement defined as the net excess number of galaxies within a
projected radius of 0.5\,Mpc (at the cluster redshift) of the central galaxy
with magnitudes between $m_1$ and $m_1 + 3$, $m_1$ being the magnitude of the
brightest cluster galaxy\footnote{For the quasars, the value of $m_1$ was
estimated from the redshift using the Hubble K$-z$ relation for radio galaxies
of the same radio power (e.g. Lilly \& Longair 1984, Best \etal\ 1998); this
is appropriate due to the similarity of radio galaxy and quasar hosts at high
redshifts (e.g. Lehnert \etal\ 1992, Kukula \etal\ 2001).}.
\nocite{lil84a,bes98d,leh92,kuk01} The value of $N_{0.5}$ was calculated for
each field, in each case using the region of the image at projected distances
greater than 1\,Mpc distant from the AGN to estimate the background counts,
thus accounting to a large an extent as possible for the cosmic variance in
the background sources.

The values of $N_{0.5}$ derived are provided in Table~\ref{cntsind}.  Although
the errors on each individual value are large, it is notable that the three
values significantly above zero correspond to the three fields with large
excesses in their number counts. This implies that the number count excesses
are predominantly in the inner regions around the AGN, and that these are
therefore consistent with cluster environments. Hill and Lilly (and see also
Wold \etal\ 2000\nocite{wol00}) calculated a conversion between the value of
$N_{0.5}$ and the traditional measure of Abell richness. $N_{0.5} \sim 10$
corresponds to the an Abell richness 0 cluster, whilst $N_{0.5} \sim 20$
corresponds to Abell richness 1. Three of the 6 fields studied are compatible
with being relatively rich clusters, whilst the other three seem to be at most
of Abell class 0 richness. It should be noted, however, that conclusions based
upon galaxy counts alone are not robust.

\subsection{Galaxy colours}

The presence of red galaxies associated with a cluster ought to redden the
average galaxy colours with respect to blank--field expectations, and indeed
this is seen in the mean or median colours of each field (see
Table~\ref{cntsind}).  Interestingly, the average colour in the radio galaxy
fields is typically reddest in the fields with the lowest number counts,
whereas naively the opposite result would be expected if the excess number
counts were solely associated with typically red cluster galaxies. This result
may imply that there are excess red galaxies around all of the AGN, and in the
cases where the non-cluster (foreground and background) galaxy counts are
randomly lower, the red cluster galaxies are therefore a larger proportion of
the total counts and influence the average colours more. This should offer
further caution about interpreting the cluster environments based on counts in
a single colour only.

\begin{figure*}
\centerline{
\psfig{file=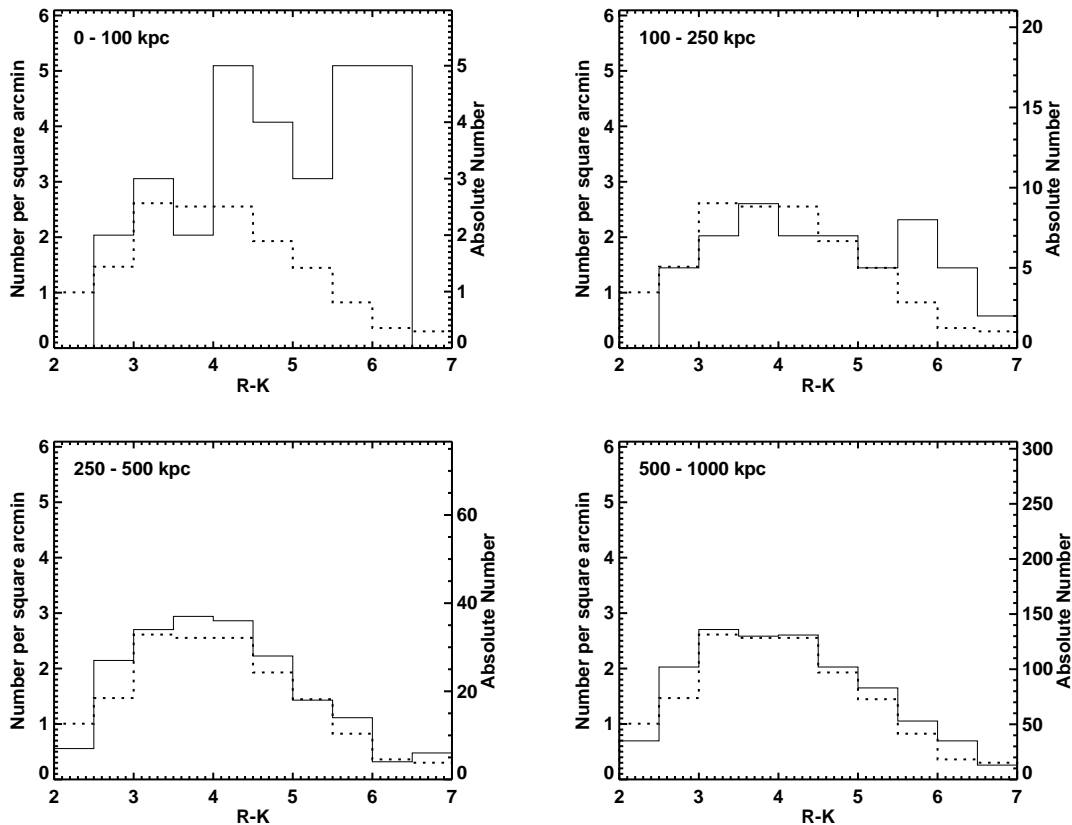,width=15cm,clip=} 
}
\caption{\label{colradhistfig} The R$-$K colour distribution of the galaxies
as a function of radius, combined over the 5 fields. The solid lines show the
colour distribution in each range of projected distances from the AGN (as
evaluated at each radio source redshift), and the dashed lines show the
distribution constructed from the regions of the images at projected radial
distances greater than 1\,Mpc. A large excess of red galaxies is seen, most
pronounced in the inner 100\,kpc but also very significant even out to
distances of a Mpc.}
\end{figure*}

To investigate in more detail the nature of the excess galaxies, the R$-$K
colour distribution of the galaxies was derived as a function of projected
distance from the AGN for all galaxies with K--magnitudes in the range $16 < K
< 21$ across the five fields (excluding 2128$-$208 for which $R$--band data is
lacking). Four radial bins were considered (0--100, 100--250, 250--500 and
500--1000\,kpc, as evaluated at the redshifts of the AGN), and the colour
distribution of the galaxies in these radial ranges are shown in
Figure~\ref{colradhistfig}. Also shown as a dotted line in these plots is the
colour distribution of those sources beyond 1\,Mpc radius from the AGN,
constructed as a comparison sample. There is a clear surplus of red ($R-K \gta
4$) galaxies at all radii $< 1$\,Mpc compared to the outer `blank--field'
regions. This is most pronounced in the inner 100\,kpc radius, but is very
significant all the way out to 1\,Mpc.

\begin{table}
\caption{\label{mediancols} A comparison of the median $R-K$ colours (together
with the errors on the mean) of galaxies in the central regions of the AGN
fields, the entirety of the AGN fields, and the blank--field distribution from
the CADIS 16hr field and K20 survey.}
\begin{tabular}{ccccccc}
K-band  & \multicolumn{6}{c}{$R-K$ Colour} \\
Magnitude& \multicolumn{2}{c}{Central Region} & \multicolumn{2}{c}{All image} & 
\multicolumn{2}{c}{Blank fields} \\
          & Median & Error & Median & Error & Median & Error \\
16--17 &  3.44  &  0.13 &  3.50  & 0.06  &  3.45  & 0.07  \\
17--18 &  3.69  &  0.14 &  3.75  & 0.07  &  3.65  & 0.05  \\
18--19 &  4.12  &  0.11 &  4.02  & 0.06  &  3.83  & 0.04  \\
19--20 &  4.20  &  0.09 &  4.08  & 0.05  &  3.86  & 0.03  \\
20--21 &  3.98  &  0.06 &  3.89  & 0.03  &  3.75  & 0.02  \\
\end{tabular}
\end{table}

\begin{figure*}
\centerline{
\psfig{file=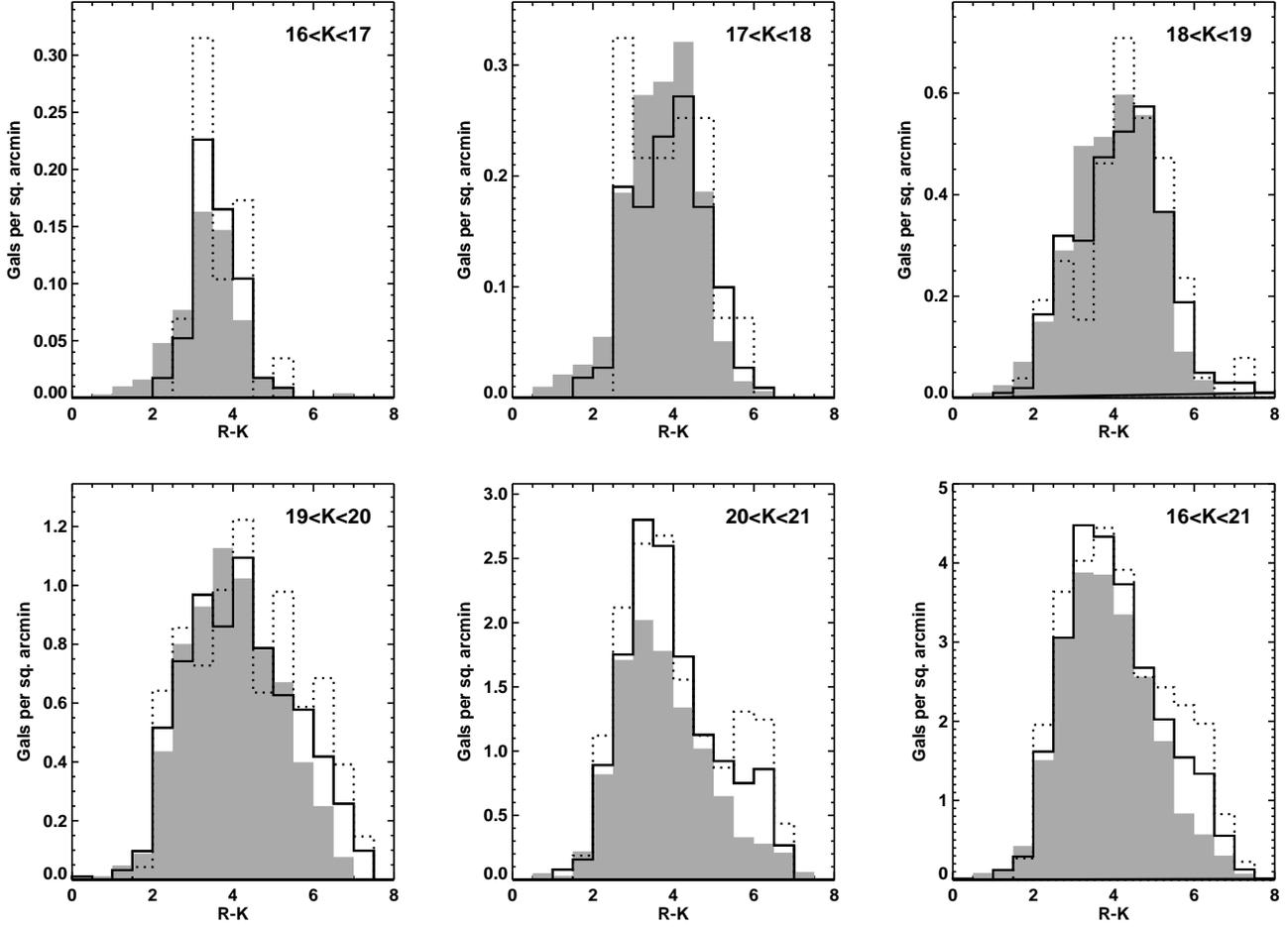,angle=90,width=\textwidth,clip=} 
}
\caption{\label{coldistfig} The $R-K$ colour distribution of the galaxies as a
function of galaxy $K$-band magnitude. The solid lines represent the
distribution combined across all of the 5 fields; the dotted line is the
distribution considering only the inner quarter of those fields, corresponding
to distances out to about 700\,kpc (at $z \sim 1.5$) from the AGN; the shaded
region represents the expected field distribution, as derived at $K<20$ by
combining the CADIS 16hr field (Thompson \etal\ 2000) and the K20 survey
(Cimatti \etal\ 2002) and for $20 < K < 21$ from the compilation of Hall \&
Green (1998).}
\end{figure*}

This colour distribution was also derived as a function of K--band magnitude,
both over the total combined area of the five images and over just the inner
quadrant of the images, corresponding to distances out to about 700\,kpc from
the AGN. These distributions are displayed in Figure~\ref{coldistfig}. The
median colours of the galaxy distributions in each magnitude range can be
found in Table~\ref{mediancols}. To compare these, a blank--field distribution
was also determined by combining the data of the CADIS 16hr field (Thompson
\etal\ 1999;\nocite{tho99} data kindly supplied by Dave Thompson) and the K20
survey (Cimatti \etal\ 2002;\nocite{cim02} data kindly supplied by Andrea
Cimatti). The data from the CADIS survey were colour corrected to account for
differences in the $R$ and $K$ passbands according to the relation $R-K =
(R-K')_{\rm CADIS} + 0.07$. The distributions of the CADIS and K20 data were
then well matched. These two fields cover a combined sky area of 206 square
arcminutes, and are complete for $K < 20$. For $20 < K < 21$, a blank--field
galaxy $R-K$ distribution was constructed using the compilation of Hall \&
Green \shortcite{hal98}, colour correcting the passbands according to $R-K =
(r-K)_{\rm HG} - 0.26$ \cite{hal98}. These expected blank--field distributions
are indicated by the shaded regions in Figure~\ref{coldistfig} and the median
colours tabulated in Table~\ref{mediancols}.

At magnitudes $K < 18$, the colour distribution of the total fields and the
central regions are both well matched to the field distribution, with
comparable median colours. By $18 < K < 19$ the median colour begins to become
redder in the AGN fields than in the blank fields, particularly in the central
regions, due to a small excess of galaxies with $R-K \gta 5$. This red excess
becomes even more pronounced in the two fainter bins. The combined colour
distribution for all galaxies with $16 < K < 21$ clearly shows that the higher
number counts in these fields are associated with red galaxies: the
blank--field predictions are matched very well for galaxies bluer than $R-K
=3$, but the AGN fields show a strong surplus of redder galaxies, particularly
those with $R-K > 5$. These results are even more pronounced when only the
central regions of the images, around the AGN, are considered.

Combining the results for the colour distributions of the galaxies as a
function of both radius and magnitude, it is clear that the predominant excess
galaxy population corresponds to a population of red ($R-K \gta 4-5$)
galaxies, with magnitudes $K \gta 18$, lying at projected distances within
$\sim 1$\,Mpc from the AGN. These are exactly the typical magnitudes, colours
and locations that would be expected of elliptical galaxies in a cluster
environment at the AGN redshifts, if such ellipticals form at early cosmic
epoch and evolve relatively passively. By contrast, in semi--analytic models
of galaxy formation, very few ellipticals this red and this luminous would be
expected in the field at redshifts $z \sim 1.6$ (cf. Daddi \etal\
2000b\nocite{dad00b}).

\nocite{tho99,cim02,hal98}

\begin{figure*}
\centerline{
\psfig{file=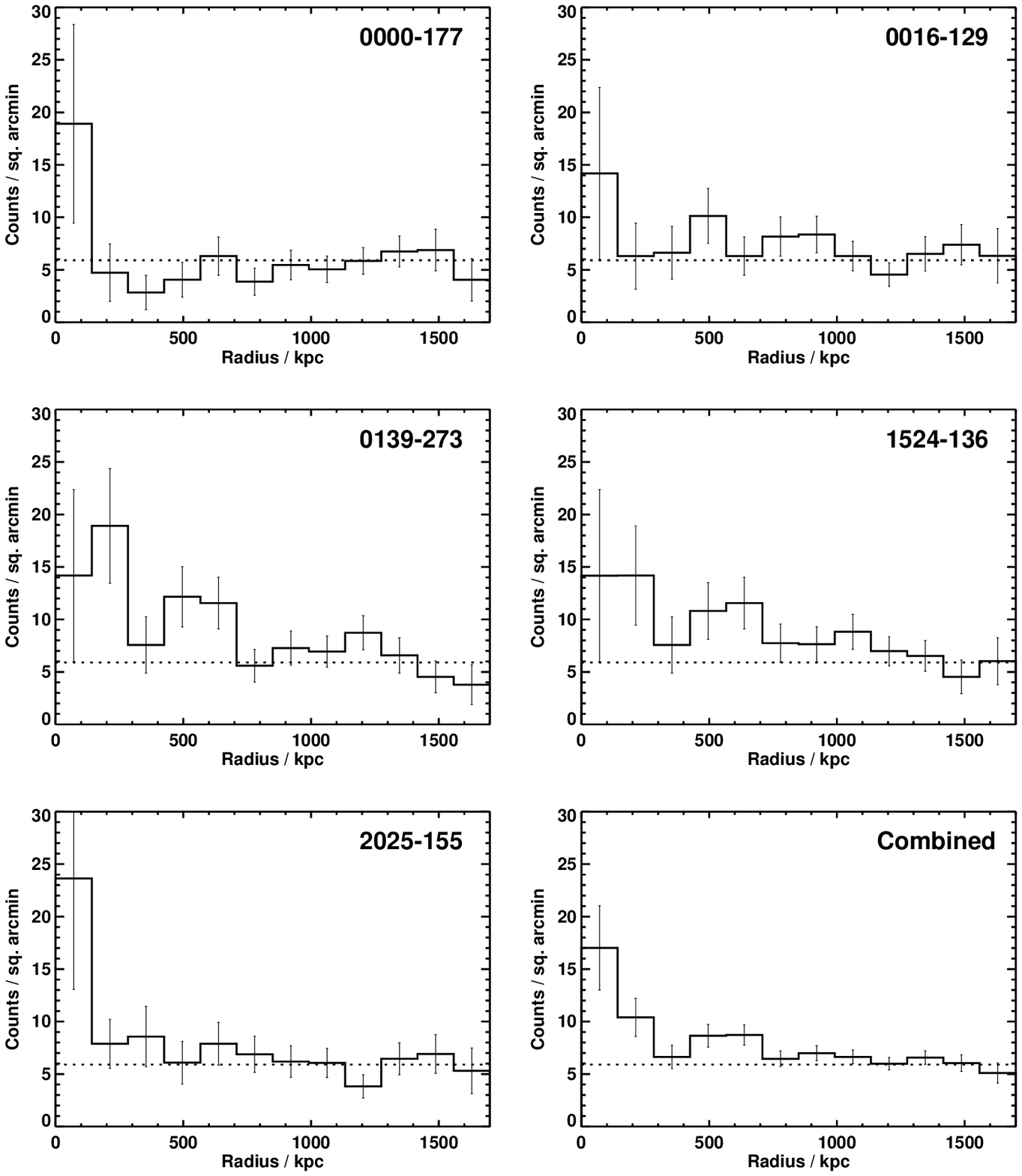,width=13.7cm,clip=} 
}
\caption{\label{redgalhists} The radial distribution of galaxies with
$17.5 < K < 20.7$ and $R-K > 4$ in each of the five fields separately, and
the average of these. The radius is calculated in kpc at the redshift of
each AGN. The error bars represent the Poissonian errors on the galaxy
counts in each bin. In each figure the horizontal dotted line represents
the expected blank--field distribution.}
\end{figure*}

\subsection{Radial distribution of individual fields}

Knowing the magnitudes and colours of the galaxies producing the excess
counts, it is possible to investigate their radial distribution in each of the
galaxy fields separately. Considering only those galaxies with magnitudes
$17.5 < K < 20.7$ and colours $R-K > 4$, the radial distribution of the galaxy
counts was derived for each field. These results are shown in
Figure~\ref{redgalhists}, together with a combined histogram produced by
averaging across all five fields. Also included on these plots is the expected
number density for blank-field sources of these magnitudes and colours; this
was calculated by taking the K20 and CADIS surveys, making the same cuts in
colour and magnitude as for the AGN fields, and then convolving the results
with the completeness versus magnitude function (Figure~\ref{compplot}).

The results from combining the five AGN fields are in agreement with those
derived by Hall \& Green \shortcite{hal98}, namely that a pronounced peak of
red galaxies is seen within the inner $\sim$150\,kpc, together with a weaker
large--scale excess. With the larger field--of--view of the data provided
here, it is now possible to quantify the radial extent of this large--scale
overdensity: the galaxy counts remain above field expectations out to between
1 and 1.5\,Mpc radius.

Comparing the five individual fields in Figure~\ref{redgalhists}, however, it
is immediately apparent that this combined result is the average of five
fields with very different properties. 0016$-$129 and 2025$-$155 do both show
radial distributions similar to the average result, with pronounced peaks in
the inner bin followed by weak red galaxy overdensities to $\sim
1$\,Mpc. 0000$-$177 has a central pronounced peak of red galaxies, but shows
no large--scale excess (indeed quite the opposite); this may be indicative of
a compact group of galaxies. 0139$-$273 and 1524$-$136 are both much less
centrally concentrated, with smoother central peaks and much more gradual
radial declines in red galaxy surface density out to the blank--field
expectations at radial distances of between 1 and 1.5\,Mpc. There is clearly a
large range in the nature, scale and significance of any galaxy overdensities
around these AGN.

That the radial distribution of galaxies within these large--scale
overdensities is not more sharply peaked should not be surprising. A galaxy
cluster with a mass of a few $\times 10^{14} M_{\odot}$ and an initial scale
of 3--5\,Mpc has an overall density of 1--4 $\times 10^{-25}$\,kg\,m$^{-3}$.
The time that such a structure will take to separate out from Hubble flow and
collapse is $t \sim (3 \pi / 32 G \rho)^{1/2} \sim$ 3--6 Gyr \cite{pea99}. The
Universe at redshifts $z \sim 1.5$ is about 4.5\,Gyr old, and so such
structures will still be collapsing into a virialised cluster such as those we
see today; the large--scale structures seen around the AGN might even not yet
be gravitationally bound.

\subsection{Bright red galaxies and sub-structure}

At high redshifts most clusters observed to date have several giant galaxies
within them, presumably due to merging subclusters. Further, there is some
evidence that AGN host galaxies are not always the only bright galaxies in
their environment, for example due to sub-cluster mergers or being part of a
group falling into a larger structure (e.g. Bremer, Baker \& Lehnert 2002;
Simpson \& Rawlings 2002)\nocite{bre02b,sim02}. Therefore, the fields have
been searched for any galaxies other than the AGN that are brighter than
$K=18$ and satisfy the colour criteria of the `cluster elliptical candidates'
above. The images around these bright red galaxies have then been examined;
the majority of these galaxies appear to be fairly isolated, and so are likely
to be foreground extremely red objects with redshifts of $z \sim 1$, since
such foreground objects will tend to dominate the extremely red galaxy
population at brighter magnitudes (e.g. Cimatti \etal\ 2002)\nocite{cim02}.
However, in two cases these bright red galaxies have associated fainter red
galaxies, and may represent substructures at the AGN redshift.

The first of these cases is in the field of 0000$-$177, where three galaxies
with $K=17.41$, $K=17.77$ and $K=17.85$ and colours $R-K>5$ are found close
together, along with fainter red galaxies, about an arcminute east of the
AGN. Indeed, these galaxies represent the small peak in the red galaxy radial
distribution (Figure~\ref{redgalhists}) at 600 to 700\,kpc radius around this
source. This region of the field is of comparable richness in red galaxies to
the regions around the AGN. The second case is a group found around a
$K=17.09$ galaxy about 1.5 arcminutes to the south--west of 0016$-$129,
partially causing the 800--1000\,kpc bump in the radial distribution around
that source. This appears even richer than the red galaxy distribution around
the radio galaxy itself, which might explain why the value of $N_{\rm 0.5}$
for this source is fairly low even though the number counts are well above the
literature average.

It is not clear whether these extra structures are indeed at the same redshifts
as the AGN, but if so then this would indicate that any large--scale structure
around the AGN is still in the process of building up. It is interesting that
this evidence for substructure is found in the two cases with the weakest
evidence for red galaxy excesses in the 200--500\,kpc radius range.

\subsection{The morphology--density relation}

In nearby clusters, a well--known morphology--density relation exists
\cite{mel77,dre80}, whereby in regions of high galaxy density, such as the
centre of the cluster, higher proportions of early--type galaxies are
found. Dressler \etal\ \shortcite{dre97} investigated the evolution of this
morphology--density relation with redshift, out to $z \sim 0.5$, and find that
this is still present at these redshifts but only in high-concentration,
regular clusters; no correlation between galaxy--type and local galaxy density
is found for low-concentration, irregular clusters. They interpret this as
meaning that the mechanisms that produce morphological segregation
(e.g. dynamical friction, or conversion of late--types into early types by gas
stripping and halting star formation) work most rapidly in high mass systems.

The nature of the small--scale galaxy excess around these distant radio sources
(and in the two potential substructures) is particularly interesting because,
where present, it is comprised almost entirely of red galaxies (cf. Figs 4 to
6).  This indicates that morphological segregation may already have taken place
in the very inner regions around the AGN, implying that the morphology--density
relation is imprinted into cluster centres at a very early epoch.

\subsection{Angular cross--correlation analyses}
\label{crosscorsec}

The clustering of galaxies around the radio galaxies can be investigated using
the angular cross--correlation function $w(\theta)$, which is defined from the
probability ($\delta P$) of finding two sources in areas $\delta\Omega_1$ and
$\delta\Omega_2$ separated by a distance $\theta$: $\delta P = N^2 [1 +
w(\theta)] \delta\Omega_1 \delta\Omega_2$, where $N$ is the mean surface
density of sources on the sky.

The value of $w(\theta)$ can be evaluated for all galaxy--galaxy pairs, for just
AGN--galaxy pairs, or restricting the galaxies included to certain ranges of
colours and magnitudes. Here, four different set of galaxy magnitudes and
colours were considered: (1) `all galaxies'; (2) `blue galaxies'; (3) `red
galaxies'; and (4) `cluster elliptical candidates', which are the subset of the
red galaxies that have K magnitudes, $R-K$ and $J-K$ colours in the range that
would be expected for old passive cluster galaxies. The exact colour and
magnitude definitions of each class of objects, and the number of galaxies
contained in that class, are provided in Table~\ref{crosstab}.  

For each of these subsamples of galaxies the value of $w(\theta)$ was
estimated for a variety of bins in $\theta$, following the method described by
B00, considering both all galaxy--galaxy pairs and only AGN--galaxy pairs. The
results obtained over all 5 fields with multi--colour data were averaged to
produce overall results, with an estimate of the uncertainty in the value of
the combined $w(\theta)$ in each bin provided by the scatter in the values
between the different fields. $w(\theta)$ usually shows a power--law form,
$w(\theta) = A (\theta / {\rm deg})^{-\delta}$, with a canonical value for the
exponent of $\delta = 0.8$; this value appears also to be appropriate at high
redshifts \cite{gia98}, and so was adopted here for all fits here. This does
not always provide an excellent fit to the data, but the current data do not
have sufficient signal--to--noise to warrant an attempt to independently fit
this parameter; note that because this parameter is fixed, the formal errors
on the cross-correlation amplitudes may be slightly higher than those quoted.

\begin{figure*}
\begin{tabular}{ccc}
{\large\bf All galaxy--galaxy pairs}
&
\hspace*{3.4cm}
&
{\large\bf AGN--galaxy pairs only}
\end{tabular} 
\centerline{
\psfig{file=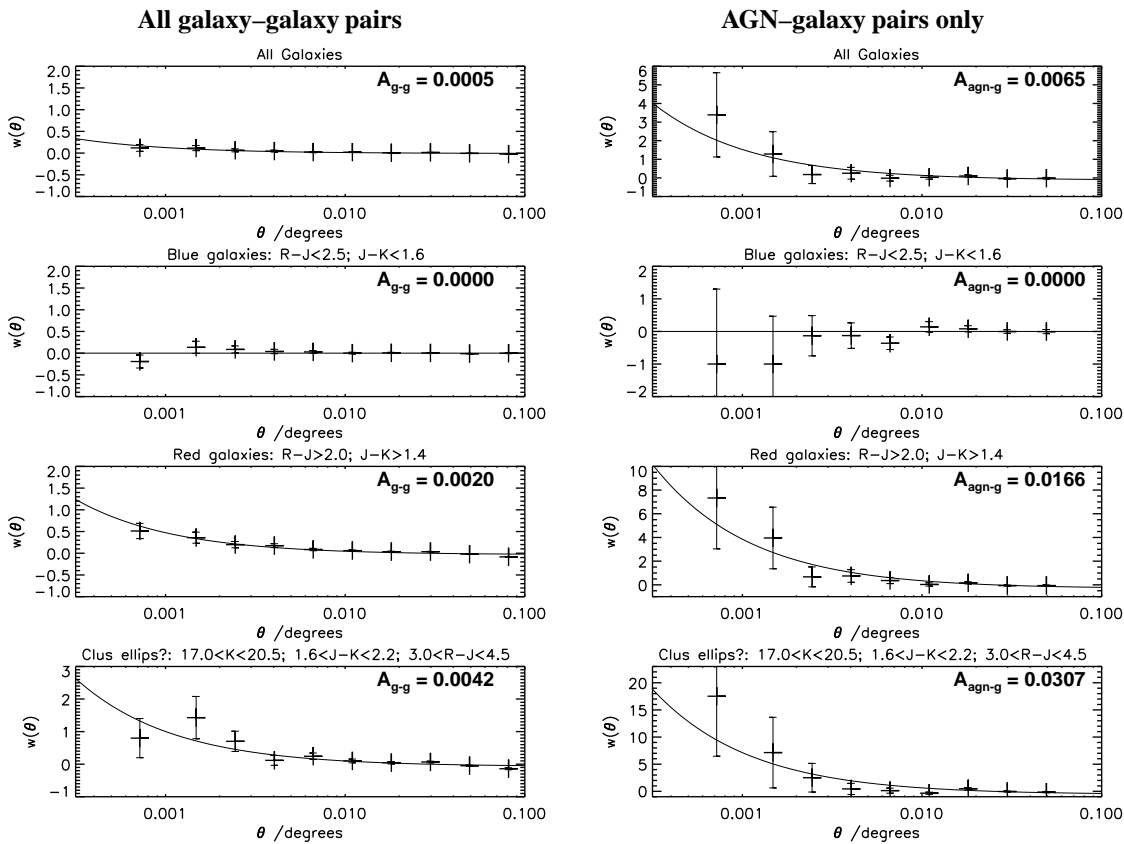,width=15cm,clip=} 
}
\caption{\label{crossfig1} The angular cross--correlation functions for
galaxies with a range of magnitudes and colours as defined in
Table~\ref{crosstab}, averaged across the 5 fields. The left--hand plots
correspond to all galaxy--galaxy pairs and the right--hand plots consider only
AGN--galaxy pairs.} 
\end{figure*}

The finite area of the images leads to a small correction, known as the
integral constraint ($C$), such that $w(\theta)_{\rm obs} = A [(\theta / {\rm
deg})^{-0.8} - C]$. The value of $C$ can be estimated by integrating
$w(\theta)$ over the area of each field, and for the current data corresponds
to $C_{\rm gg} = 16.7$ and $C_{\rm ag} \approx 19.1$ for all galaxy-galaxy
pairs and just AGN-galaxy pairs respectively. Using these expressions, the
amplitudes of the angular cross--correlation function were derived for both
all galaxy--galaxy pairs and just AGN--galaxy pairs for each of the selected
colour--magnitude subsamples of galaxies. The cross--correlation functions are
displayed in Figure~\ref{crossfig1}, and Table~\ref{crosstab} provides the
derived cross--correlation amplitudes.

\begin{table*}
\caption{\label{crosstab} Results from the angular cross correlation
analyses. For galaxies with a range of magnitudes and colours, the amplitude
of the angular cross-correlation function is provided considering both all
galaxy--galaxy pairs ($A_{\rm gg}$) and only AGN--galaxy pairs ($A_{\rm
ag}$).}
\begin{tabular}{lcccrcc}
~~~~~Sample & K--mag range& IR col range& IR-opt col range &$N_{\rm gals}$&
\multicolumn{2}{c}{Angular cross-correlation amplitude}\\
       & &     &  & & $A_{\rm gg}$ &$ A_{\rm ag}$ \\
All galaxies      & K$<$20.7    &    All      &    All       &1739 & 
$0.0005 \pm 0.0001$ & $0.0065 \pm 0.0022$\\
Blue galaxies     & K$<$20.7    & J-K$<$1.6   & R-J$<$2.5    & 762 & 
$0.0000 \pm 0.0003$ & $0.0000 \pm 0.0026$\\
Red galaxies      & K$<$20.7    & J-K$>$1.4   & R-J$>$2.0    & 868 & 
$0.0020 \pm 0.0002$ & $0.0166 \pm 0.0041$\\
Cluster ellips? &17.0$<$K$<$20.5&1.6$<$J-K$<$2.2&3.0$<$R-J$<$4.5&252&
$0.0042 \pm 0.0004$ & $0.0307 \pm 0.0052$\\
\end{tabular}
\end{table*}

The angular cross--correlation amplitude of `all galaxies' in these fields is
within the range of the values recently derived to $K=20.75$ for
`blank--fields' (e.g. Carlberg \etal\ 1997; Roche \etal\ 2002; Daddi \etal\
2000a)\nocite{roc02,car97c,dad00a}. However, comparing the values of $A_{\rm
gg}$ and $A_{\rm ag}$ in Table~\ref{crosstab} shows that the galaxies are much
more highly clustered around the AGN than around other galaxies in general,
indicating that a significant proportion of the galaxies are connected with
the AGN. Evidence for such clustering around the AGN is particularly apparent
when comparing the results for galaxies of different colours. The blue
galaxies show no evidence for clustering, not even around the AGN, whilst red
galaxies are known to be more highly clustered (e.g. Roche \etal\ 1996,
Shepherd \etal\ 2001 and references therein)\nocite{roc96,she01} and this is
reflected in the value of $A_{\rm gg}$ derived for this class; their higher
value of $A_{\rm ag}$ shows that these are preferentially clustered around the
AGN. The `cluster elliptical candidates' have even larger values of $A_{\rm
gg}$ and $A_{\rm ag}$, suggesting that this selection criteria picks out a
highly clustered class of object, many of which are indeed likely to be old
passive cluster ellipticals. It should be noted that this clustering signal is
dominated by the points at small radii, $\theta \lta 10$\,arcsec,
corresponding to the excess in the upper left panel of Figure 12; however,
even if the inner two bins of the cross--correlation analysis are ignored, the
clustering amplitude around the AGN is still higher than that of the field in
general, indicating that it is not only these nearby galaxies which produce
the clustering signal.

If a population of $N_{\rm c}$ cluster galaxies, with a high intrinsic angular
cross--correlation amplitude ($A_{\rm c}$), is observed together with a sample
of $N_{\rm i}$ foreground or background interloper galaxies with zero clustering
amplitude, then the resultant combined sample will have an observed
cross--correlation amplitude of $\approx \frac{N_{\rm c}}{N_{\rm c} + N_{\rm i}}
A_{\rm c}$. In other words, the observed amplitude will be the amplitude that
would be found for the cluster galaxies alone, scaled down by the proportion of
the total sample that are cluster galaxies. This situation is a good
approximation to those of the `red galaxy' and `cluster elliptical candidate'
samples, since the observed value of $A_{\rm gg}$ for these samples are almost
an order of magnitude below those of $A_{\rm ag}$. Comparing the values of
$A_{\rm ag}$ then indicates that the colour definition for the `cluster
elliptical candidate' sample approximately halves the fraction of interlopers
relative to the red galaxies.

If the results from the number count analysis are taken at face value, then
there are about 270 clusters galaxies across the five fields studied in the
cross--correlation analysis (290 excess galaxies determined in
Section~\ref{subgalcnts}, of which 20 are from the 2128$-$208 field, not
considered here); using this, an estimate can be made for the number of
cluster galaxies in each of the colour--selected samples.  Assuming that about
80\% of the cluster galaxies are `red' (from Figure~\ref{coldistfig}) then
within the `red galaxy' sample there are about 220 cluster galaxies, which
corresponds to about 25\% of the overall `red galaxy' sample (see
Table~\ref{crosstab}). This would suggest that about 40\% of the `cluster
elliptical candidate' sample, that is $\sim$100 galaxies, are indeed cluster
galaxies. If these results are correct, the $\sim$100 early--type cluster
galaxies at redshifts $z \sim 1.6$ would provide a substantial sample for
cluster galaxy evolution studies.

As has been described by many authors (e.g. Longair \& Seldner 1979, Prestage
and Peacock 1988)\nocite{lon79b,pre88}, if the form of the galaxy luminosity
function at the redshift of the AGN is assumed, it is possible to convert the
amplitude of the angular cross--correlation function for galaxies around the
AGN into a spatial cross--correlation amplitude, $B_{\rm ag}$, derived from
the spatial cross--correlation function $\xi(r) = B_{\rm ag}
\left(\frac{r}{{\rm Mpc}}\right)^{-\gamma}$. Following the method of B00, the
value of $A_{\rm ag} = 0.0061 \pm 0.0022$ for the `all galaxies' sample
corresponds to a spatial cross--correlation amplitude of $B_{\rm ag} = 660 \pm
240$. This amplitude is comparable to the value of $B_{\rm ag} = 510 \pm 120$
derived by B00 around 3CR radio galaxies with redshifts $z \sim 1$. These
values can be interpreted physically by comparing with the equivalent values
($B_{\rm cg}$) for Abell clusters calculated between the central galaxy and
the surrounding galaxies; these have been derived by several authors and
average to $B_{\rm cg} \approx 350$ for Abell class 0 and $B_{\rm cg} \approx
710$ for Abell class 1 \cite{pre88,hil91,and94,yee99}.  The environments
surrounding these $z \sim 1.6$ AGN seem to be, on average, comparable to those
of local clusters of richness between Abell classes 0 and 1.

\section{Environmental variations compared with radio source properties}
\label{envcomp}

The results have clearly shown that the environments of these high redshift
AGN are heterogeneous, with considerable variation in both the amplitude and
the physical extent of any galaxy excess. In this section the variations in
the galaxy overdensities are compared with the other properties of the radio
source, to see if there are any indications of why the environments differ so
much. The properties of the radio sources studied are provided in
Table~\ref{sampdet}. 

The first thing to note is that all of these radio sources have essentially
the same redshifts and radio powers, and so the variations are not generally
correlated with either of these properties. Equally, all 6 studied sources are
steep--spectrum radio sources according to the standard definition of $\alpha
> 0.5$ (where $\alpha$ is defined as $S_\nu \propto \nu^{-\alpha}$), although
1524$-$136 has a spectral index close to this cut--off; the weaker
small--scale excess of this source could conceivable be related to this in
some way. No fundamental difference is seen between the quasar and radio
galaxy populations, and nor is any correlation with the size of the radio
source seen for any of the extended radio sources.

The only issue which this comparison does bring to light is that the one
source for which no overdensity is seen on either large or small scales,
2128$-$208, is an unresolved radio source; this could potentially explain this
result. Extended radio sources must have been fueled continuously for
timescales of at least $10^6$ years, and possibly up to $10^8$ years, which
requires a plentiful and steady supply of fueling gas, as well as a relatively
dense surrounding medium to confine the expanding radio lobes. Perhaps at high
redshifts these conditions are only satisfied for galaxies within
larger--scale structures. On the other hand, compact radio sources are
relatively young, $\lta 10^5$ years old (e.g. Owsianik and Conway
1998)\nocite{ows98b}, and it has been suggested that some proportion of these
may fizzle out before evolving into larger radio sources (e.g. Alexander
2000).\nocite{ale00} The reduced gas supply required for these short--lived
compact sources may then be available to galaxies in any environment, and
could explain the lack of any overdensities around 2128$-$208. 

This situation has been modelled by Kauffmann \& Haehnelt \shortcite{kau02}
who have investigated the quasar--galaxy cross--correlation function around
quasars at different redshifts within hierarchical clustering models of galaxy
formation. In their model they assume that quasars are triggered by galaxy
mergers, with some percentage of the gas brought in by the merger then
fuelling the black hole. They show that the cross--correlation function around
the quasars is significantly higher for quasars with long lifetimes than for
short--lived sources, in agreement with the discussion above. This difference
is most pronounced for the most powerful AGN, such as those studied here. 

An alternative may be that this is a lower power radio source with its flux
density Doppler boosted by beaming; in this case a correlation between radio
power and environmental density would explain the result. However, a beaming
scenario is considered unlikely due to the steep spectral index of the source.

\begin{table*}
\caption{\label{sampdet} Properties of the 6 radio sources studied.}
\begin{tabular}{lcccccccc}
~~~~~Source & z & $S_{\rm 408}$ &$\alpha_{1400}^{408}$ & 
$P_{\rm 408}$ & D  & Type & \multicolumn{2}{c}{~~Excess on scale$^*$}\\
&&[Jy]&&[$10^{27}$\,W\,Hz$^{-1}$]&[$''$]& & $\lta$150\,kpc & $\sim$1\,Mpc\\ 
\\
MRC 0000-177 & 1.47 & 6.51 & 0.80 & 85.4 & 2.7 & Q  & Y & N \\ 
MRC 0016-129 & 1.59 & 6.87 & 0.95 &125.4 & 3.5 & RG & Y & y \\
MRC 0139-273 & 1.44 & 5.04 & 0.95 & 72.2 & 12  & RG & y & Y \\
MRC 1524-136 & 1.69 & 6.11 & 0.61 & 92.5 & 0.4 & Q  & y?& Y \\ 
MRC 2025-155 & 1.50 & 5.41 & 1.05 & 93.9 & 15  & Q  & Y & y \\
MRC 2128-208 & 1.62 & 6.15 & 0.88 &110.1 & $<1$& Q  & N & N \\ 
\\
\end{tabular}
\\
$^*$ 'Y' = strong-excess; 'y' = weak excess; N = no significant excess.
\end{table*}

\section{Discussion and Conclusions}
\label{discuss}

The results of this work can be summarised as follows.

\begin{itemize}
\item On average, a significant overdensity of K--band selected galaxies with
$17.5 < K < 20.5$ is found across the AGN fields.

\item These excess counts are predominantly associated with galaxies with 
magnitudes $K \gta 17.5$ and colours $R-K \gta 4$, which are (although not
uniquely) exactly those expected for old passive cluster galaxies at redshifts
$z \sim 1.6$.

\item These galaxies are found on two different spatial scales around the AGN:
pronounced concentrations of galaxies at radial distances $\lta 150$\,kpc, and
weaker large--scale overdensities extending out to between 1 and 1.5\,Mpc.

\item The presence or absence of galaxy excesses on these two scales
differs greatly from AGN field to AGN field. All of the extended radio sources
show overdensities on some scale: two show overdensities on both scales, two
predominantly on large scales and one only on small scales.  The one field
which shows little evidence for a galaxy excess on any scale is associated
with an unresolved radio source.

\item Where overdensities are present on $\lta 150$\,kpc scales, these are
composed almost entirely of red galaxies. This suggests that the
morphology--density relation is imprinted into the centres of clusters at a
very early stage of cluster formation.

\item The amplitude of the angular cross--correlation function around the AGN
is a strong function of galaxy colour. The selection of only galaxies with
magnitudes and colours consistent with being old passive elliptical galaxies
at the AGN redshifts maximises this amplitude, indicating that old ellipticals
do indeed exist in these environments.
 
\item The spatial cross--correlation amplitude for all galaxies around the AGN
is similar to that around the 3CR radio galaxies at redshifts $z \sim 1$; both
are consistent with having the same richness as local Abell clusters of richness
class 0 to 1.
\end{itemize}

\noindent It is interesting to consider what these results imply for the
nature of the AGN at these redshifts, and also for cluster evolution.

Powerful radio galaxies are known to be hosted by giant elliptical galaxies at
all redshifts $z \lta 1.3$ (e.g. Best \etal\ 1998, McLure \& Dunlop
2000)\nocite{bes98d,mcl00a}, and even out to redshifts $z < 2.4$ a significant
proportion still have radial profiles well--matched by de Vaucouleurs' law
\cite{pen01}. Little difference is found between the host galaxies of radio
galaxies and radio--loud quasars at these redshifts \cite{kuk01}, which
suggests that the hosts of all of the $z \sim 1.5$ AGN studied here are likely
to be giant elliptical galaxies.

The variety of different environments found for these AGN may simply reflect
this fact: that the primary requirement for forming a powerful radio--loud AGN
is a giant elliptical galaxy host. Such galaxies can be found in a range of
environments from galaxy clusters, through groups of galaxies, to relatively
isolated environments, but are generally only formed by redshifts $z \sim 1.6$
when they are found at the highest peaks of the primordial density
perturbations, within larger--scale structures. This explains the high average
richness, but still significant variation, of the large--scale galactic
environments around these AGN.

To understand the pronounced smaller scale galaxy overdensity, the key
question is whether these central galaxies represent a stable virialised
structure, or whether they are still infalling and will merge with the central
radio galaxy. If these are already virialised then the small scale galaxy
overdensity is clearly related to the larger--scale structure, and indicates
that at least some of the AGN lie at a special location in their environments:
at the centres of their clusters or groups. This again is not too surprising:
the kinetic energy of the relativistic radio jets of these AGN corresponds to
the Eddington luminosity of a black hole with $M \sim 10^9\,M_{\odot}$
\cite{raw91b}, and so even assuming that these sources are fueled close to the
Eddington limit they must be powered by very massive central engines and the
radio lobes must be confined by dense gas. It is not unreasonable to expect
these most powerful radio sources to be found within the most massive
elliptical galaxies, at the centre of the cluster or group of galaxies.

In this interpretation, the four different scenarios of the presence or absence
of galaxy overdensities on small and large scales correspond to four different
possible locations of the AGN host galaxy: it may be a central cluster galaxy in
cluster with a virialised core (excesses on both scales), an off--centre giant
elliptical in a cluster, or the central galaxy in a cluster at an early stage of
formation (only large--scale overdensity), the central galaxy of a compact group
(only small--scale excess), or an isolated giant elliptical (no excess on any
scale). With respect to this model, it is interesting that all of the {\it
extended} radio sources lie in group or cluster environments.

Alternatively, if the galaxies producing the small--scale excess are not yet
virialised, then they may instead be related to a hierarchical build up of the
central AGN host galaxy. Pentericci \etal\ \shortcite{pen01} found that the
optical characteristic sizes of powerful radio galaxies are typically a factor
of two smaller at $z \sim 2$ than those at $z \sim 1$. Since the galaxy
luminosity and the characteristic size of ellipticals are related by $L_{\rm
int} \propto r_{\rm e}^{0.7}$ \cite{kor77}, this means that the radio galaxies
must grow in mass by about a factor of two over this redshift interval, a
large part of which may be due to mergers with companion galaxies. If the
150\,kpc scale galaxy excess has a velocity dispersion $\sigma \sim
500$\,km\,s$^{-1}$, then the galaxy crossing time for this structure is $\sim
0.3$\,Gyr. A galaxy merger timescale of a few crossing times is therefore
comparable to the interval of cosmic time between redshifts $z \sim 2$ and $z
\sim 1$, and some of the small--scale excesses of galaxies may be destined to 
merge with the AGN host galaxy soon. 

Whichever of these interpretations for the small--scale galaxy excess around
the AGN is correct (and undoubtedly a combination of both effects is
occurring), the Mpc--scale galaxy overdensities appear be associated with
(proto?) cluster environments. The wealth of galaxies with magnitudes and
colours consistent with being old elliptical galaxies at $z \sim 1.6$ indicate
that the epoch of elliptical galaxy formation must be earlier than this
redshift; the colours of these objects further suggest that the bulk of their
stellar populations must have formed at redshifts $z \gta 5$ (this will be
discussed further in Best \etal\ in preparation). By using these radio--loud
AGN as signposts, it is therefore possible to define significant samples of $z
\sim 1.6$ cluster early--type galaxies with which to extend studies of galaxy
evolution back to earlier cosmic epochs.

Finally, it is interesting to compare the comoving number density of these
powerful radio sources with that of nearby rich clusters. According to the
pure luminosity evolution model for steep--spectrum radio sources of Dunlop
and Peacock \shortcite{dun90}, at redshifts $z=1$ to 2, there are about
$10^{-7}$ radio sources per comoving Mpc$^3$ within the upper decade of radio
luminosities containing the sources studied here. By comparison, the comoving
number density of nearby clusters of Abell richness class 0 or greater is
$\sim 4 \times 10^{-6}$ Mpc$^{-3}$, and for Abell class 1 and richer it is
$\sim 1.7 \times 10^{-6}$ Mpc$^{-3}$ (e.g. Bahcall \& Soniera 1983; converted
to the cosmology adopted in this paper; cf. West 1994).\nocite{bah83,wes94}
However, radio sources only live for between $10^7$ and $10^8$ years
(e.g. Kaiser \etal\ 1997),\nocite{kai97b} and so the observed number density
of powerful radio sources underestimates the number density of environments
which will contain a powerful radio source at some point between redshifts 2
and 1 by a factor $\sim 50$ ($=$ cosmic time between redshift 2 and 1 divided
by the lifetime of a typical radio source). Therefore, the comoving number
density of environments which will contain a powerful radio source between
redshifts 2 and 1 is comparable to the comoving number density of nearby Abell
clusters of richness class 0 or greater. This may simply be an interesting
coincidence, or alternatively may imply that the onset of a powerful radio
source is a key stage in the formation of a rich cluster.

\section*{Acknowledgements} 

PNB would like to thank the Royal Society for generous financial support through
its University Research Fellowship scheme. The authors are very grateful to Dave
Thompson and Andrea Cimatti for providing data on the R-K distributions of the
CADIS and K20 surveys in electronic form. We thank the referee, Malcolm Bremer,
for a careful reading of the original manuscript and helpful comments. This work
is based upon observations made at the European Southern Observatory, La Silla,
Chile, proposals ESO:65.O-0590(A,B) and ESO:67.A-0510(A,B). This research has
made use of the NASA/IPAC Extragalactic Database (NED) which is operated by the
Jet Propulsion Laboratory, California Institute of Technology, under contract
with the National Aeronautics and Space Administration. This research has made
use of the USNOFS Image and Catalogue Archive operated by the United States
Naval Observatory, Flagstaff Station. This work was supported in part by the
European Community Research and Training Network `The Physics of the
Intergalactic Medium'.

\label{lastpage}
\bibliography{pnb} 
\bibliographystyle{mn} 

\end{document}